\documentclass[10pt,letterpaper]{article}
\usepackage{multicol} % Used for the two-column layout of the document
\usepackage[letterpaper,top=1in,bottom=1in,left=1in,right=1in]{geometry}
\usepackage{times}
\usepackage{graphicx}
\usepackage{url}
\usepackage{hyperref}

\title{TCP SYN COOKIE VULNERABILITY}
\author{Dakshil Shah, Varshali Kumar\\ \{shah329, kumar261\}@purdue.edu
}
\date{May 3, 2018 }

\begin{document}
\maketitle

\begin{abstract}
TCP SYN Cookies were implemented to mitigate against DoS attacks. It ensured that the server did not have to store any information for half-open connections. A SYN cookie contains all information required by the server to know the request is valid. However, the usage of these cookies introduces a vulnerability that allows an attacker to guess the initial sequence number and use that to spoof a connection or plant false logs.\\
\end{abstract}

\begin{multicols}{2} % Two-column layout throughout the main article text 

%Include your sections here. Introduction and Conclusion are necessary.
\section{Introduction} The idea behind SYN cookies was to process an ACK packet without the need to store any information[1]. Everything required to validate an ACK packet is encoded within the Initial Sequence Number generated by the server. This is what we call a SYN Cookie. It contains the following : \begin{enumerate}
\item A timer value that increments every 64 seconds (5 bits)
\item An encoding of an MSS selected by the server in response to the client's MSS (3 bits)
\item A secret function, selected by the server, of the Client IP, client port , server IP, server port and timer value. (24 bits)
\end{enumerate}

\section{Literature Survey}
The TCP protocol was made to a be a reliable transport layer protocol. It required that unacknowledged data packets  be re-transmitted. This is where the issue arises when using cookies. In normal mode, when the server sends SYN-ACK but does not receive the corresponding ACK it would re-transmit.\\ 
Eventually it would send a reset (RST) to the client to shut it down. However if cookies are in use, the server would not know that the returning ACK packet is lost and retransmission would thus not be possible. \\

In a study by Lemon (2002), he explains the problems with using TCP syn-cookies. He observed that a TCP connection established in cookie mode depends entirely on the final ACK packet.\\ 
The final ACK is completely independent of the initial SYN and SYN-ACK packets. This implies an attacker can simply bombard a server with ACK packets containing random values of ISN, out of which hopefully, one would be correct.\\ 
This would allow a connection to be established. Using only ACKs would allow an attacker to even bypass firewalls, since most of them filter out incoming packets with SYN bit set.\\

\section{The Vulnerability}
\includegraphics[scale=0.8]{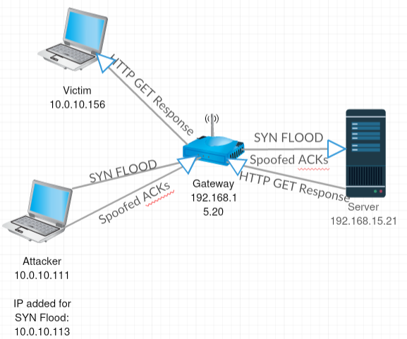}\\
\centerline{Fig 1: Network Setup}\\

With syn-cookies there is a reduced cost of guessing the ISN for a successful connection forgery. Currently, the kernel accepts cookies that were created less than 2 minutes ago. This gives 2 delta counter values (0-1) that are accepted by the server.\\ It accepts 4 MSS values. In total, server would accept \(2*4 = 8\), combinations of timer t and MSS values. Each of these would give a correct ISN with the secret function.\\
To an attacker, guessing the value is now 8 in \(2^{32}\). Initially, the kernel accepted 4 counter values and 8 MSS values. Even though the number of valid ISNs has been reduced from 32 to 8, we show that a connection forgery is still feasible. \\

We used VMs to avoid bringing down an actual network or server. The victim does no activity and just exists on the network. The server runs an apache server on port 80 with a few files hosted. The attacker and victim are on the same network and can talk to the server via the Gateway.\\

Before we carry out the attack, the target server must be in cookie mode. However, current implementations use TCP-SYN-ACK mode with a backlog queue.\\

\noindent \textit{Backlog queue size:\\ cat /proc/sys/net/ipv4/tcp\_max\_syn\_backlog}\\

Only when the backlog overflows i.e. in the case of a SYN flood and having the backlog filled with half open connections, does the kernel start using SYN cookies.\\

\noindent \textit{check number of half open connections:\\ netstat -an | grep -c SYN\_RECV} \\

\includegraphics[scale=0.4]{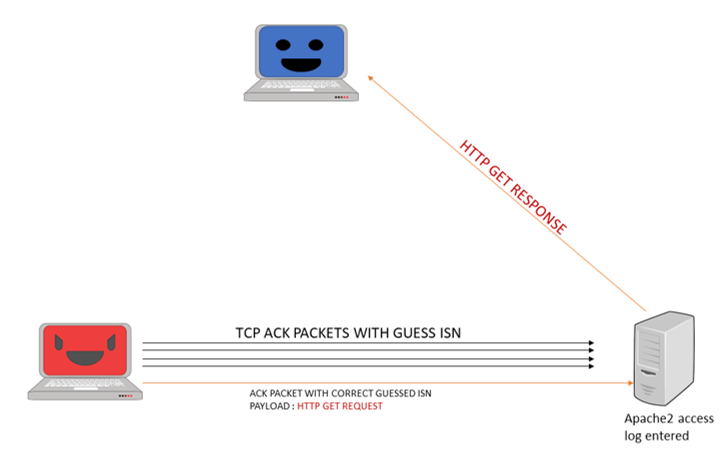}\\
\centerline{Fig 2: The Attack}\\

To do this we use our attacker to first carry out a TCP-SYN flood, forcing the server into using cookies.\\ 
On the attacker machine, we add another IP address which drops all incoming packets. We do this as if we send a SYN packet from this address, the server will send a reset packet after some time or the network sends a host unavailable. To prevent this, we add the address and drop incoming packets.\\

Our aim is to forge a connection from a spoofed IP to a server with a GET request. If it is successful the server would have false logs for a file download. Since the server does not remember any state, we can forge connection by guessing ISN in ACK packets.We crafted ACK packets with guessed ISNs and the payload was a GET request.\\

\section{Results}

\includegraphics[scale=0.4]{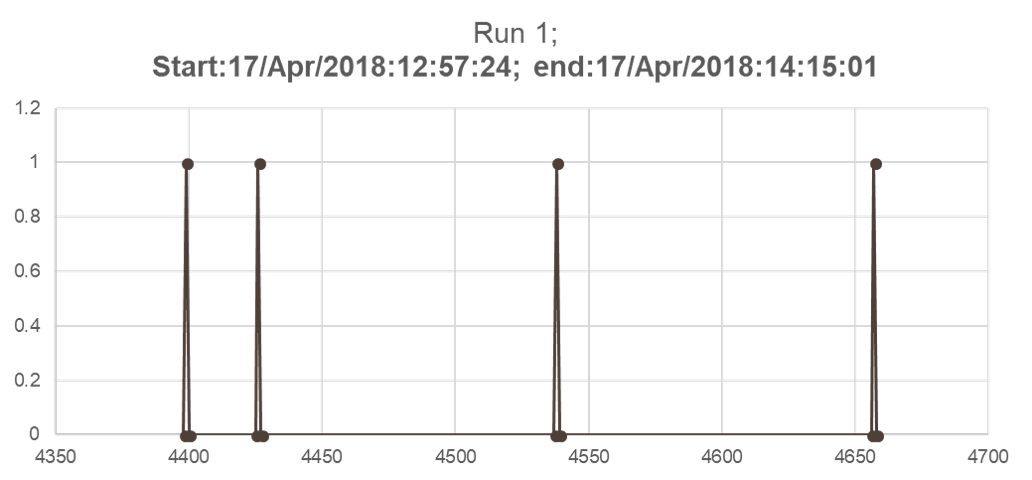}\\
\centerline{Fig 3: Virtual Machine Run 1}\\

A linear search is currently not possible as it is infeasible to send all possible packets before the counter value changes. Thus a large integer is used to cover the space uniformly. We increment the initial ACK number using this large integer rather than a linear increment.\\

We observed that if the initial ACK number is carefully chosen wrt to the time of the day and the servers response to a test query, the attack can be optimized to run faster.\\

\includegraphics[scale=0.4]{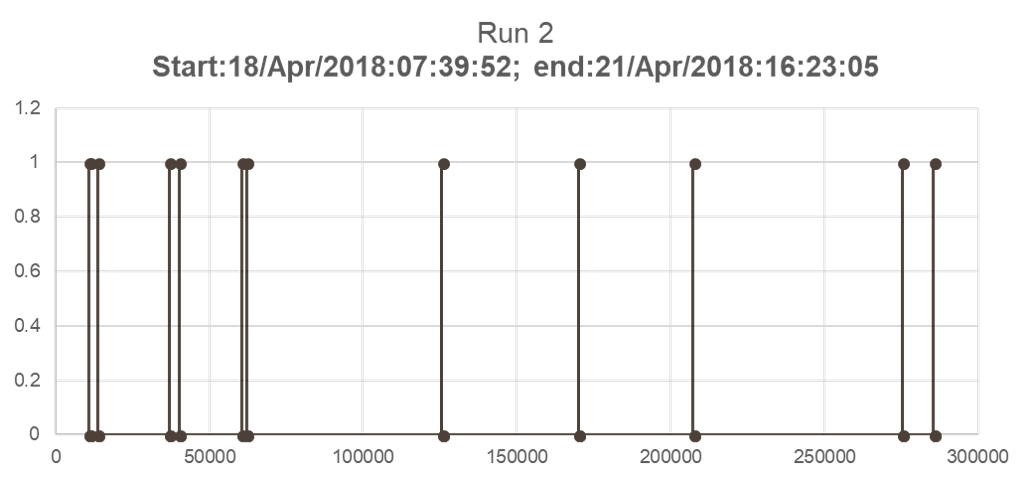}\\
\centerline{Fig 4: Virtual Machine Run 2}\\

\includegraphics[scale=0.4]{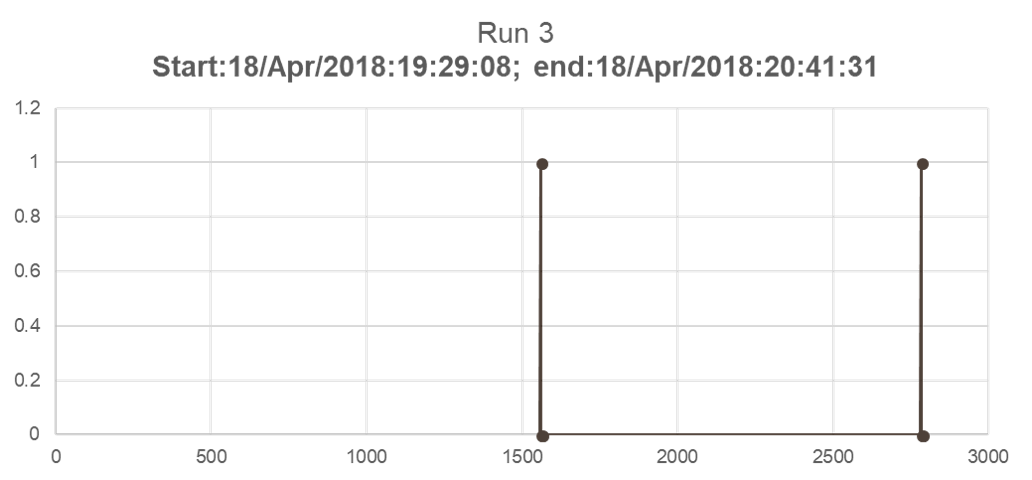}\\
\centerline{Fig 5: Real Machine Run}\\

As the attack depends in various factors such as when the attack was started, the initial ACK number and increment set, we observed attacks at random times, with the best being within 26 minutes of start, and also some within a minute of each other.\\

\noindent The graphs show that the attack is spaced over time. Certain gaps occur in the virtual machine runs due to the servers on which the VMs were hosted, randomly killed/paused long running jobs(here the VM instances)

\noindent The results are got from the server logs. As there are too many packets being sent, along with normal network traffic, using tcpdump to capture was not possible. The tcpdump buffer overflows and discards packets, due to which the results also end up getting lost. We inspected the log of the apache server. The log contains a record only if the connection was successfully completed ie spoofed.\\

\section{Conclusion}

With networks getting faster, it would be possible to send more packets in a minute to cover more possible valid combinations for the SYN cookie. This reduces the time needed for the attack and makes it more feasible to be carried out in a public setting such as a coffee shop. The attack demonstrated also shows that we do not need to guess all 32 bits of the ISN. When SYN cookies are in use, it is sufficient to guess the last 24 bits with all possible valid combinations of the leading 8 bits.\\

Currently, our version of the attack is not possible in a public setting without further optimization. In our trials on real devices over a wireless network, it led to the network crashing, preventing others from accessing
the Internet. In the virtual machine environment setup on a single machine and also on real machines connected directly via LAN cables, the attack was successful.\\

\section{Next Steps}

While a good intrusion detection system should be able to detect a SYN flood, this may lead to a false assumption that the attackers intention is to DOS the server. This may preemptively lead to the system being put into SYN Cookie mode which is exactly what is desired. An IDS or Network admin should consider the possibility that they might be simply being used to frame someone or have fake access logs planted.

Consider an attack scenario where an attacker wants to access some files on the system. This may be an environment where classified data is stored and only insiders have access to it. To cover ones tracks, an insider may download the classified files but at the same time carry out this attack. Now the log files are no longer untainted. Audit logs to be admissible in court as legal evidence have a requirement that they were not tampered with. Following the attack, it is difficult to successfully prove that the defendant did indeed download the files. Their information may have simply been planted during the attack.[5]\\

While this vulnerability is known, the scope for misuse has not been considered. An attacker may already know how to use this exploit to cover their tracks or even frame an innocent by spoofing activity to questionable sites. Due to this, a public disclosure is needed rather than a private disclosure. Since it's very difficult to modify the TCP protocol to increase the ISN field beyond 32 bits[1].\\

The only feasible mitigation currently possible is to block all requests from an IP, which come within microseconds of each other. However, if the ip is spoofed, the legitimate user with that IP would also get blocked.\\

{\raggedright
\small
\bibliographystyle{unsrt}

}

\appendix
%If you want to include something in appendix

\end{multicols}	
\end{document}